\documentclass[12pt]{article}

\usepackage{graphicx}
\usepackage{float}
\usepackage{colordvi}
\usepackage{graphicx}
\usepackage{amsmath,amsfonts,amssymb}

\usepackage{authblk}

\title{Measures of uncertainty in market network analysis}

\author[1]{Kalyagin V.A.}
\author[2]{Koldanov A.P.}
\author[2]{Koldanov P.A.}
\author[1,3]{Pardalos P.M.}
\author[1]{Zamaraev V.A.}

\affil[1]{National Research University Higher School of Economics, Laboratory of Algorithms and
Technologies for Network Analysis, Nizhny Novgorod, Russia}
\affil[2]{National Research University Higher School of Economics, Department of Applied Mathematics and Informatics, Nizhny Novgorod, Russia}
\affil[3]{Center for Applied Optimization, University of Florida, Gainesville, FL, USA}

\begin{document}

\maketitle



\begin{abstract}
    Statistical uncertainty of different filtration techniques for market network analysis is studied. Two measures of statistical
    uncertainty are discussed. One is based on conditional risk for multiple decision statistical procedures and another one is based on average
    fraction of errors. It is shown that for some important  cases the second measure is a particular case of the first one.
    Statistical uncertainty for some popular market network structures is analyzed. Results of numerical evaluation of statistical uncertainty for
    minimum spanning tree, market graph, maximum cliques and maximum independent sets are given. The most stable structures are derived.
\end{abstract}


\section{Introduction}

In this paper we study uncertainty in market network analysis.
Network models of financial markets attract a growing attention
last decades \cite{Mantegna99, TAMM05, BCLMVM04, BBP05},
especially in Econophysics \cite{ManStan, Maman, MamanGan, MBLM,SJY, NJR, WXCYY}.
A common network representation of the stock market is based on Pearson correlations of return
fluctuations.
In such a representation each stock corresponds to a vertex and a link between two
vertices is estimated by sample Pearson correlation of corresponding returns. The obtained network is complete weighted graph. In order to
simplify the network and preserve the key information different filtering techniques are used in the
literature.

One of filtering procedures is the extraction of a minimal set of important links associated with the highest
degree of similarity belonging to the minimum spanning tree (MST) \cite{Mantegna99}. To construct the MST,
a list of edges is sorted in descending order according to the weight and following the ordered list
an edge is added to the MST if  and only if it does not create a cycle. The MST was used to
find a topological arrangement of stocks traded in a financial market which has associated a meaningful
economic taxonomy. This topology is useful in the theoretical description
of financial markets and in search of economic common factors affecting specific groups of stocks. The
topology and the hierarchical structure associated to it, is obtained by using information present in the
time series of stock prices only.

The reduction to a minimal skeleton of links leads to loss of valuable information. To overcome this issue in
\cite{TAMM05} it was proposed to extent the MST by iteratively connecting the most similar nodes until the
graph can be embedded on a surface of a given genus $g = k$. For example, for $g = 0$ the resulting graph
is planar, which is called Planar Maximally Filtered Graph (PMFG). It was concluded in \cite{TAMM05} that
the method is pretty efficient in filtering relevant information about the connection structure both of the whole system and within obtained clusters.

Another filtering procedure leads to the concept of market graph
\cite{BBP03, BBP05, BBP06}. A market graph (MG) is obtained from
the original network by removing all edges with weights less than
a specified threshold $\theta \in [-1,1]$. Maximum cliques (MC)
and maximum independent sets (MIS) analysis of the market graph
was used to obtain valuable knowledge about the structure of the
stock market. For example, it was noted in \cite{VGKKK13} that the
peculiarity of the Russian market is reflected by the strong
connection between the volume of stocks and the structure of
maximum cliques. The core set of stocks of maximum cliques for
Russian market is composed by the most valuable stocks. These
stocks account for more than 90\%  of the market value and
represent the largest Russian international companies from banking
and natural resource sectors. In contrast, the core set of stocks
of the maximum cliques for the US stock market has a different
structure without connection to the stocks values.

Today network analysis of financial market is a very active area
of investigation and various directions are developed in order to
obtain valuable information for different stock markets
\cite{HZY09, WXCYY, MG, VGKKK13, WXCC13, BKKKP, NGUYEN}. Most of
these approaches use time series observations. It is well known
that financial time series have a stochastic nature. Therefore,
any such analysis needs to be complemented by estimation of
statistical uncertainty of obtained results. The main question is:
how reliable are the conclusions regarding market structures such
as MST, PMFG, MG, MC, MIS and others? In the other word, does
market network analysis lack of confidence?

To answer this question we propose two measures of statistical uncertainty of market network analysis.
Both measures of uncertainty
in our approach are a number of observations needed to obtain results with given confidence.
To measure statistical uncertainty we use conditional risk for multiple decision statistical procedures \cite{KKKP13} and PFE-type error 
(per-family error rate \cite{HochTamh}) which we call {\it fraction of error}.
We show by simulations that MG, MC, MIS are more stable with respect to statistical uncertainty than MST and PMFG.
In particular for 250 stocks of US market to obtain 90\% of confidence for MST one needs at least 10000 observations,
 but the same confidence for MG is already reached with 300 observations.

The paper is organized as follows.
In Section \ref{sec:NetMatAnalysis} we introduce notations and definitions of market network
analysis used in the paper.
In Section \ref{sec:statUncert} we describe two approaches to measure the statistical uncertainty of market structures.
In Section \ref{sec:expResults} we apply our method to analyze the uncertainty of US market
network.
In Section \ref{sec:note} we discuss a condition on the minimal number of observations.
In Section \ref{sec:Conclusions} we give concluding remarks.

\section{Network market analysis} \label{sec:NetMatAnalysis}
All graphs in this paper are simple, i.e. finite, undirected, without loops or multiple edges.
We represent a stock market as a weighted graph $G=(V,E)$, where $V = \{1,2, \ldots, N\}$ is the set of
vertices (stocks) and $s_{ij}$ is the weight (similarity measure between stocks) of the edge $(i,j) \in E$.

For a connected graph, a spanning tree is an acyclic connected subgraph
which contains all the vertices.
Minimum spanning tree (MST) of a connected, weighted graph is a spanning tree of this graph of the
minimal possible weight, where a weight of a tree is the sum of the weights of the edges in the tree.
MST of a given graph may be constructed iteratively by choosing among the edges of the graph not yet chosen
the edge of the smallest weight which does not form any cycle with those edges already chosen. This
method is known as Kruskal's algorithm \cite{K56}.
Note that in market network analysis $s_{ij}$ represents similarity between stocks $i$ and $j$. In this
setting we are interested in a spanning tree of maximal weight (maximum spanning tree), which can be
found by the same algorithm by choice on each step of the edge of the largest weight. We will still refer to it
as minimum spanning tree (MST).
Planar maximally filtered graph (PMFG) is constructed in direct analogy with MST where an edge with
the largest weight is added under the constraint that on each step of the algorithm the obtained graph is planar \cite{TAMM05}.

A market graph (MG) is defined as a graph in which vertices correspond to stocks and two vertices
are connected by an edge if and only if a similarity measure between the corresponding stocks exceeds
a specified threshold $\theta$ \cite{BBP03}.
In a graph, a clique is a subset of pairwise adjacent vertices and an independent set is a subset of vertices
no two of which are adjacent. Maximum clique (MC) is a clique of maximal size.
Maximum independent set (MIS) is an independent set of maximal size.
\\

\noindent\textbf{Example 1} \\
Let us consider the next 10 stocks of US market: A (Agilent Technologies Inc), AA (Alcoa Inc),
AAP (Advance Auto Parts Inc), AAPL (Apple Inc), AAWW (Atlas Air Worldwide Holdings Inc), ABAX (Abaxis Inc),
ABD (ACCO Brands Corp), ABG (Asbury Automotive Group Inc), ACWI (iShares MSCI ACWI Index Fund), ADX (Adams Express Company).

Let $s_{ij}$ be the Pearson correlations between stocks returns calculated from time series of observations for one year period starting from November 2010. The weighting matrix is shown in the table below.

{\scriptsize
$$
    \left(
        \begin{array} {rcccccccccc} \label{arr:weightMatrix}
        \vline     &\text{A}  &\text{AA}  &\text{AAP} &\text{AAPL}&\text{AAWW} &\text{ABAX} &\text{ABD} &\text{ABG} &\text{ACWI}& \text{ADX} \\
               \hline
        \text{A}    \vline&    1.0000&    0.7220&    0.4681&    0.4809&    0.6209&    0.5380&    0.6252&    0.6285&    0.7786&    0.7909\\
        \text{AA}   \vline&    0.7220&    1.0000&    0.4395&    0.5979&    0.6381&    0.5725&    0.6666&    0.6266&    0.8583&    0.8640\\
        \text{AAP}  \vline&    0.4681&    0.4395&    1.0000&    0.3432&    0.3468&    0.2740&    0.4090&    0.4016&    0.4615&    0.4832\\
        \text{AAPL} \vline&    0.4809&    0.5979&    0.3432&    1.0000&    0.4518&    0.4460&    0.4635&    0.4940&    0.6447&    0.6601\\
        \text{AAWW} \vline&    0.6209&    0.6381&    0.3468&    0.4518&    1.0000&    0.5640&    0.5994&    0.5369&    0.7170&    0.7136\\
        \text{ABAX} \vline&    0.5380&    0.5725&    0.2740&    0.4460&    0.5640&    1.0000&    0.4969&    0.4775&    0.6439&    0.6242\\
        \text{ABD}  \vline&    0.6252&    0.6666&    0.4090&    0.4635&    0.5994&    0.4969&    1.0000&    0.6098&    0.7161&    0.7158\\
        \text{ABG}  \vline&    0.6285&    0.6266&    0.4016&    0.4940&    0.5369&    0.4775&    0.6098&    1.0000&    0.6805&    0.6748\\
        \text{ACWI} \vline&    0.7786&    0.8583&    0.4615&    0.6447&    0.7170&    0.6439&    0.7161&    0.6805&    1.0000&    0.9523\\
        \text{ADX}  \vline&    0.7909&    0.8640&    0.4832&    0.6601&    0.7136&    0.6242&    0.7158&    0.6748&    0.9523&    1.0000
        \end{array}
    \right)
$$
}

\begin{figure}[H]
    \centering
    \newlength\figureheight
    \newlength\figurewidth
    \setlength\figureheight{10cm}
    \setlength\figurewidth{16cm}
    \includegraphics[scale=1]{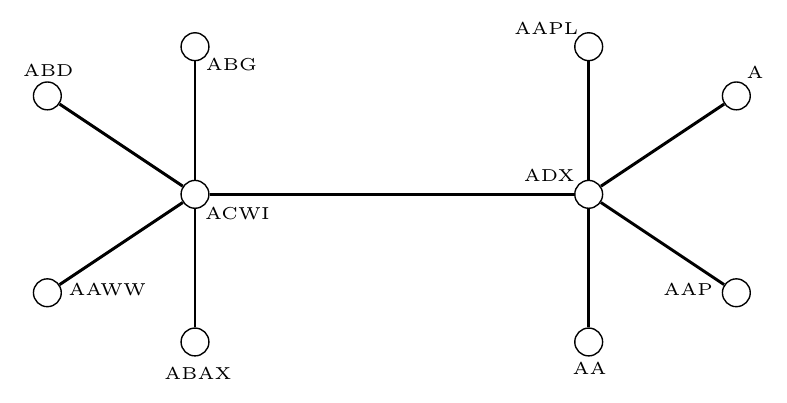}
    \caption{MST}
    \label{fig:ref_mst}
\end{figure}

\begin{figure}[H]
    \centering
    \setlength\figureheight{10cm}
    \setlength\figurewidth{16cm}
    \includegraphics[scale=1]{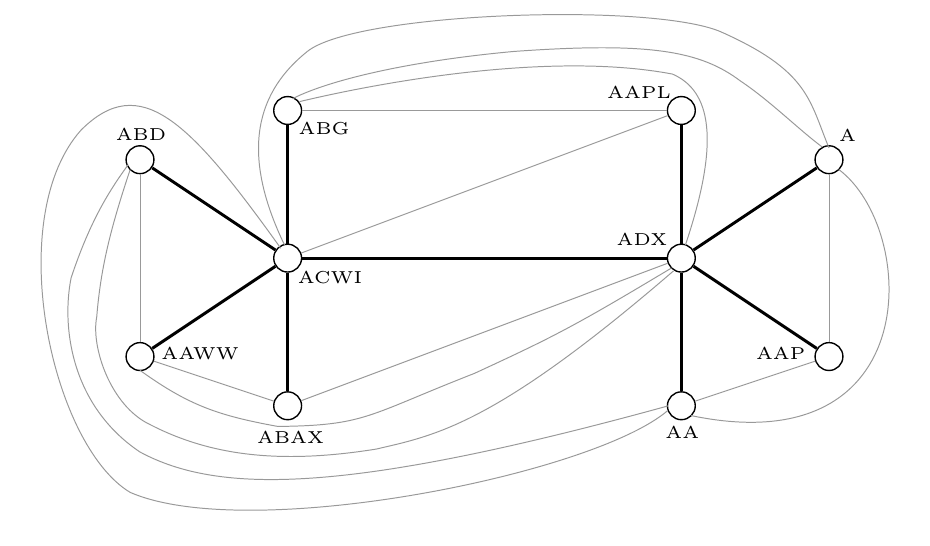}
    \caption{PMFG}
    \label{fig:ref_pmfg}
\end{figure}

\begin{figure}[H]
    \centering
    \setlength\figureheight{10cm}
    \setlength\figurewidth{16cm}
    \includegraphics[scale=1]{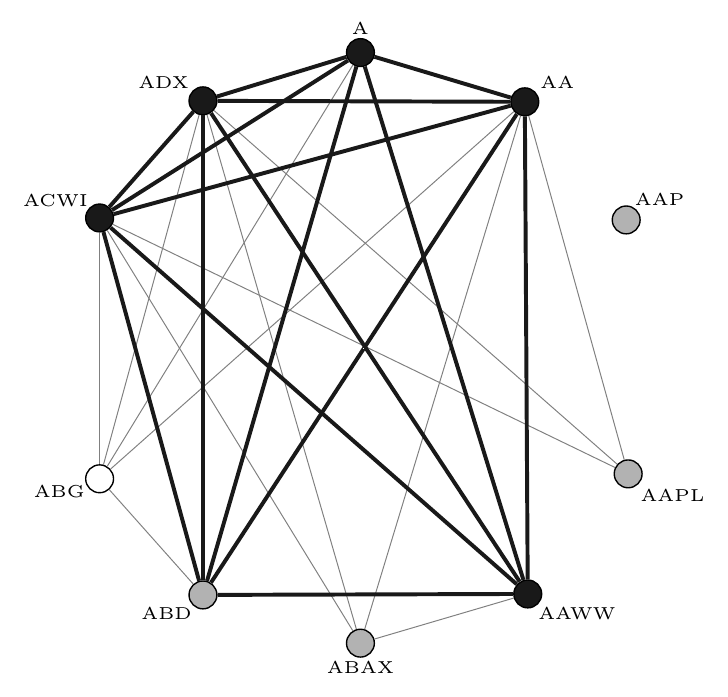}
    \caption{Market graph}
    \label{fig:ref_market_graph}
\end{figure}

The corresponding network structures are presented in Figures \ref{fig:ref_mst}, \ref{fig:ref_pmfg}, \ref{fig:ref_market_graph}.

For MST  one can observe two clusters (\{AAWW, ABAX, ABD, ABG, ACWI\} and
\{A, AA, AAP, AAPL, ADX\}) with the centers ACWI and ADX connected by the edge.

The structure of PMFG is more complicated than MST. The filtered graph contains more information
about the structure of considered market because new connections appear. There are 5 new intracluster
and 10 intercluster connections. The PMFG contains the MST, but does not have clear cluster structure.

The market graph is constructed for the threshold $\theta = 0.55$. The graph contains two
connected components one of which is almost complete graph. There are two maximum cliques with
6 vertices. One of them \{A, AA, AAWW, ABD, ACWI, ADX\} has the maximal weight.
There are four maximum independent sets with 4 vertices. One of them \{AAP, AAPL, ABAX, ABD\}
has minimal weight.

Note that the described structures (MST, PMFG, MG, MC, MIS) are unweighted subgraphs of the network and
reflect different aspects of dependence in the stock market. It is expected that analysis of these structures may
give a better understanding of the market behaviour.

\section{Measures of statistical uncertainty} \label{sec:statUncert}

Let $N$ be a number of stocks, $n$ be a number of days of observations.
In our study financial instruments are characterized by daily returns of the stocks. Stock $k$ return for day
$t$ is defined as
\begin{equation}
    R_k(t) = \ln \frac{P_k(t)}{P_k(t-1)},
\end{equation}
where $P_k(t)$ is the price of stock $k$ on day $t$.
We assume that for fixed $k$, $R_k(t)$, $t=1,\ldots, n$, are independent random
variables with the same distribution as $R_k$ (i.i.d.) and the random vector $R=(R_1, \ldots, R_N)$
has multivariate normal distribution $R \sim N((a_1, \ldots, a_N), ||\sigma_{ij}||)$,
where
\begin{equation}
    ||\sigma_{ij}|| = \left(
        \begin{array}{ccc}
            \sigma_{11} & \cdots & \sigma_{1N} \\
            \cdots & \cdots & \cdots \\
            \sigma_{N1} & \cdots & \sigma_{NN}
        \end{array}
        \right)
\end{equation}
is the covariance matrix.
For this model we introduce the \textit{reference network} which is a complete weighted graph
with $N$ nodes and weight matrix $||\sigma_{ij}||$. For the reference network one can consider corresponding
reference structures, e.g. reference MST, reference PMFG, reference market graph and others.

Let $r_k(t), k=1,\dots, N, t=1, \ldots, n$, be the observed values of returns. Define the \textit{sample covariance}
$$
    s_{ij} = \frac{1}{n-1}\sum\limits_{t=1}^{n}{(r_i(t)-\overline{r}_i)(r_j(t)-\overline{r}_j)},
$$
and {\it sample correlation} 
$$
r_{i,j}=\frac{s_{i,j}}{\sqrt{s_{i,i}s_{j,j}}}
$$
where $\overline{r}_i = \frac{1}{n}\sum\limits_{t=1}^{n}r_i(t)$.
Using the sample covariances we introduce the (\textit{$n$-period}) \textit{sample network} which is a complete weighted graph with $N$ nodes and weight matrix $||s_{ij}||$.
For the sample network one can consider the corresponding
sample structures, e.g. sample MST, sample PMFG, sample market graph and others.
Note that $(\overline{r}_1, \ldots, \overline{r}_N)$ and $\{s_{ij}| i,j = 1, \ldots, N\}$ are sufficient
statistics.

\subsection{Conditional risk} \label{subsec:measureStatUncertStat}
To handle statistical uncertainty we propose to compare the sample
network with the reference network. Our comparison will be based
on conditional risk connected with possible losses. The associated loss function is defined following  \cite{KKKP13} withing the framework of multiple
decision theory. 

For a given structure
$\mathcal{S}$ we introduce a set of hypothesis:
\begin{itemize}
    \item $h_{ij}$: edge between vertices $i$ and $j$ is not included in the reference structure $\mathcal{S}$;
           \item $k_{ij}$: edge between vertices $i$ and $j$ is included in the reference structure $\mathcal{S}$.
\end{itemize}
To measure the losses we consider
two types of errors:
\begin{enumerate}
    \item[] \textbf{Type I error:} edge is included in the sample structure when it is absent in the reference structure;
    \item[] \textbf{Type II error:} edge is not included in the sample structure when it is present in the reference structure.
\end{enumerate}


Let $a_{ij}$ be the loss associated with the error of the first kind and $b_{ij}$ the loss associated with the error of
the second kind for the edge $(i,j)$. According to the statistical decision theory \cite{Wald} and
taking into account additivity of the loss function \cite{Lehmann} we define the conditional risk for a given structure $\mathcal{S}$ as
\begin{equation}\label{eq:Rsn}
    \mathcal{R}(\mathcal{S},n) = \sum\limits_{1 \leq i < j \leq N} [a_{ij}P_n(d_{ k_{ij} }|h_{ij}) + b_{ij}P_n(d_{ h_{ij} }|k_{ij})],
\end{equation}
where $P_n(d_{k_{ij}}|h_{ij})$ is the probability of rejecting hypothesis $h_{ij}$ when it is true and
$P_n(d_{h_{ij}}|k_{ij})$ is the probability of accepting hypothesis $h_{ij}$ when it is false.
Note that conditional risk is a decreasing function on the number of observations. We say that structure
$\mathcal{S}_1$ \textit{is more stable than structure} $\mathcal{S}_2$ if
$\mathcal{R}(\mathcal{S}_1,n) < \mathcal{R}(\mathcal{S}_2,n)$ for any number of observations $n$.
In other words statistical uncertainty of structure $\mathcal{S}_1$ is less than statistical uncertainty of
structure $\mathcal{S}_2$ if $\mathcal{R}(\mathcal{S}_1,n_1) = \mathcal{R}(\mathcal{S}_2,n_2)$
implies $n_1 < n_2$.
We define the \textit{$\mathcal{R}$-measure of statistical uncertainty of structure $\mathcal{S}$ (of level $\mathcal{R}_0$)}
as the number of observations $n_{\mathcal{R}}$ such that $\mathcal{R}(\mathcal{S},n_{\mathcal{R}}) = \mathcal{R}_0$, where $\mathcal{R}_0$ is given value.

\subsection{Error rate (fraction of error).} \label{subsec:simpleMeasureStatUncert}
Another approach to measure statistical uncertainty can be based
on Per-Family Error Rate \cite{HochTamh}. In our case
errors can be defined using a number of edges different in sample
structure with respect to reference structure and vice versa.

Let
$$
    x_{1}^{ij} = \left\{
    \begin{array}{rl}
        1, & \text{if edge }(i,j) \text{ is incorrectly included into sample structure,} \\
                     0, & \text{otherwise,}
    \end{array}
    \right.
$$
and
$$
    x_{2}^{ij} = \left\{
    \begin{array}{rl}
        1, & \text{if edge }(i,j) \text{ is incorrectly not included into sample structure,} \\
                     0, & \text{otherwise.}
    \end{array}
    \right.
$$
Define
$$
    X_1 = \sum\limits_{1 \leq i<j \leq N}{x_1^{ij}}; ~~~ X_2 = \sum\limits_{1 \leq i<j \leq N}{x_2^{ij}}.
$$
$X_1$ is the number of incorrectly included edges into the sample
structure and $X_2$ is the number of incorrectly not included
edges into the sample structure. Note that $E(X_1)$ and $E(X_2)$ are similar to widely used in multiple decision theory per-family type error rate \cite{HochTamh}. 

Let us now define the random variable
\begin{equation}\label{eq:X}
    X = \frac{1}{2}\left(\frac{X_1}{M_1} + \frac{X_2}{M_2}\right),
\end{equation}
where $M_1$ -- is a maximal possible value of $X_1$ and $M_2$ --
is a maximal possible value of $X_2$. Random variable $X \in
[0,1]$ describes total fraction of errors. We define the
\textit{$\mathcal{E}$-measure of statistical uncertainty of
structure $\mathcal{S}$ (of level $\mathcal{E}_0$)} as the number
of observations $n_{\mathcal{E}}$ such that
$\mathcal{E}(\mathcal{S},n_{\mathcal{E}}) = \mathcal{E}_0$, where
$\mathcal{E}_0$ is given value and $\mathcal{E}(\mathcal{S},n) =
E(X)$.

Let us pay attention to the case when $\mathcal{R}$-measure and $\mathcal{E}$-measure coincide.
Namely, if $M_1$ and $M_2$ are constants (non random) then
$$
    \mathcal{E}(\mathcal{S},n) = \sum\limits_{1 \leq i < j \leq N} \left[\frac{1}{2M_1}P_n(x^{ij}_1 = 1) + \frac{1}{2M_2}P_n(x^{ij}_2 = 1)\right].
$$
Now if  in (\ref{eq:Rsn}) we put $a_{ij} = 1/2M_1$ and $b_{ij} = 1/2M_2$ then
$$
    \mathcal{R}(\mathcal{S},n) = \sum\limits_{1 \leq i < j \leq N} \left[\frac{1}{2M_1}P_n(d_{ k_{ij} }|h_{ij}) + \frac{1}{2M_2}P_n(d_{ h_{ij} }|k_{ij}) \right].
$$
Since $P_n(d_{ k_{ij} }|h_{ij}) = P_n(x^{ij}_1 = 1)$ and $P_n(d_{ h_{ij} }|k_{ij}) = P_n(x^{ij}_2 = 1)$,
it implies that
$$
    \mathcal{E}(\mathcal{S},n) = \mathcal{R}(\mathcal{S},n) \text{ and } n_{\mathcal{E}} = n_{\mathcal{R}},
$$
that is both introduced measures of statistical uncertainty are
equivalent. The latter fact allows us to introduce the intuitive
notion of statistical uncertainty based on
$\mathcal{E}(\mathcal{S},n)$ into the framework of general theory
of statistical decision functions \cite{Wald}. The described above relation 
between $\mathcal{R}$-measure and $\mathcal{E}$-measure can take
place for such structures as MG, MST, PMFG and it is not possible
for MC and MIS.


\section{Experimental study of uncertainty} \label{sec:expResults}


Let us consider the correlation matrix $||\rho_{ij}^{US}||$ consisting of
pairwise correlations of daily returns of a set of 250 financial instruments
traded in the US stock markets over a period of 365 consecutive trading days in 2010-2011.
We use the matrix $||\rho_{ij}^{US}||$ as a weight matrix for our reference network.
We will refer to it as the \textit{US reference network}.
To construct the $n$-period sample network we simulate the sample
$x_{11}, \ldots, x_{1N}, \ldots, x_{n1}, \ldots, x_{nN}$
from multivariate normal distribution $N((0, \ldots, 0)$, $||\rho_{ij}^{US}||), i,j =  \overline{1,N}$, $N = 250$.

In our experiments we use measure of statistical uncertainty based on $\mathcal{E}(\mathcal{S},n)$ which is much easy to estimate. 
We estimate $\mathcal{E}(\mathcal{S},n)$  in the following way:
\begin{enumerate}
    \item In the US reference network find reference structure $\mathcal{S}$.
    \item Simulate sample $x_{11}, \ldots, x_{1N}, \ldots, x_{n1}, \ldots, x_{nN}$.
           \item Calculate estimations $r_{ij}$ of parameters $\rho_{ij}^{US}$.
    \item In sample network (with weight matrix $||r_{ij}||$) find sample structure $\mathcal{S}$.
           \item Calculate fraction of errors of type I ($X_1/M_1$), fraction of errors of type II ($X_2/M_2$) and total fraction of errors ($X$) by (\ref{eq:X}).
    \item Repeat 1000 times steps 1-5 and calculate mean of $X$ which is an estimation of $\mathcal{E}(\mathcal{S},n)$.
\end{enumerate}

In our experiments we choose a level of statistical uncertainty $\mathcal{E}_0=0.1$.

\subsection{Statistical uncertainty of MST}
Observe that $X=0$ means that reference MST
and sample MST are equal; $X=1$ means that
reference MST and sample MST are completely different, i.e. have no common edges. The latter situation may
hold for several sample MSTs under fixed reference MST.
For minimum spanning tree, one has $M_1 = M_2 = N-1$, where $N$ is a number of vertices in considered network.
For each edge $(i,j)$ such that $x_1^{ij}=1$ there is an edge $(k,s)$ with $x_2^{ks}=1$ and vice versa.
It means that in MST a number of errors of type I is equal to a number of errors of type II, i.e. $X_1 = X_2$.
Since $M_1$ and $M_2$ are
constants, both measures of statistical uncertainty for MST are equivalent and can be defined from the equation:
$$
    \frac{1}{2(N-1)}\sum\limits_{1 \leq i < j \leq N} [P_n(x^{ij}_1 = 1) + P_n(x^{ij}_2 = 1)] =
    \frac{1}{(N-1)}\sum\limits_{1 \leq i < j \leq N} P_n(x^{ij}_1 = 1) = \mathcal{E}_0.
$$



\begin{figure}[H]
    \centering
    \setlength\figureheight{10cm}
    \setlength\figurewidth{16cm}
    \includegraphics[scale=0.46]{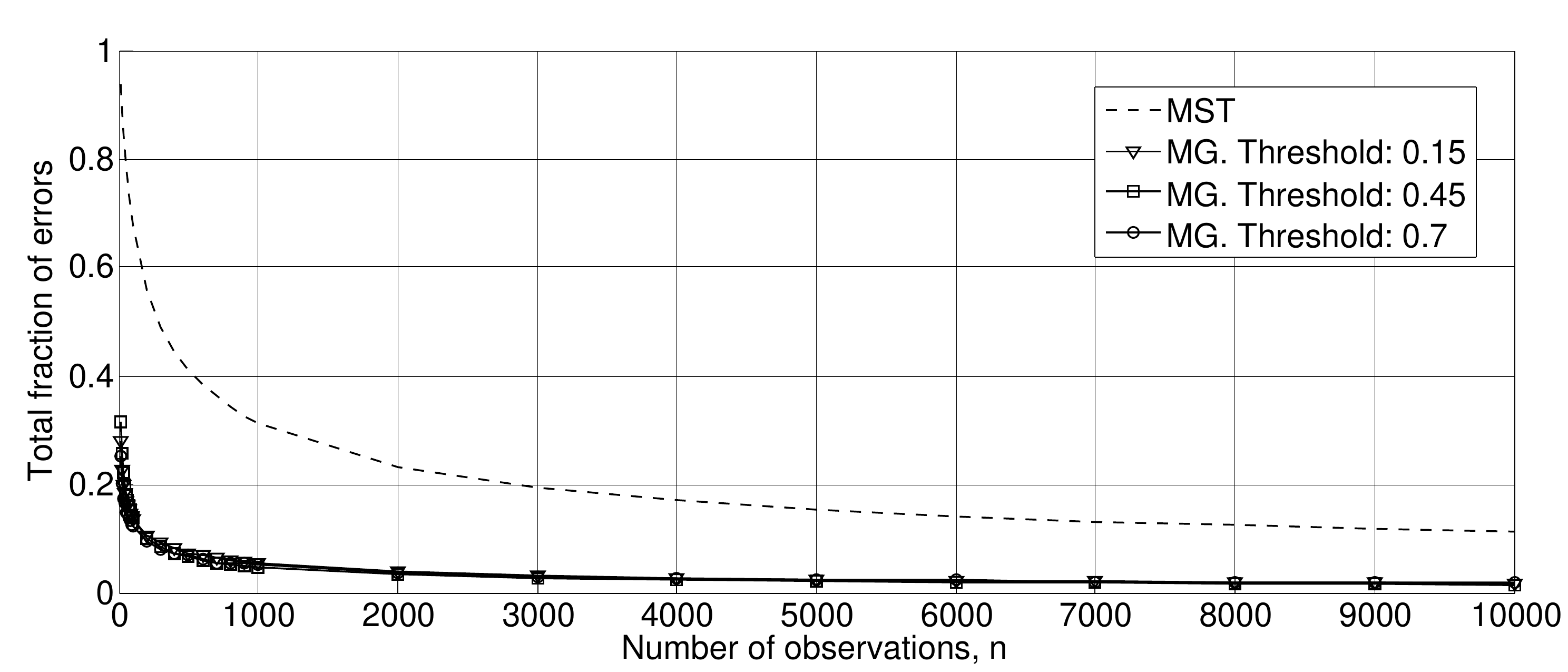}
    \caption{Total fraction of errors in MST and MG.}
    \label{fig:mst_250}
\end{figure}

Results of the study of statistical uncertainty of MST are presented in Fig. \ref{fig:mst_250}. As one can see the
condition $\mathcal{E}(\text{MST},n) \leq 0.1$ is achieved when the number of observed periods $n_{\mathcal{E}}$ is
more than 10 000.
Note that when $n=1000$ (which corresponds to 4 years of daily observations) sample and reference MSTs have only 70\% of common edges.
Moreover by further increasing the number of observations does not lead to considerable
decrease of statistical uncertainty of MST. As it was mentioned in \cite{TAMM05}, statistical uncertainty of PMFG
is higher than statistical uncertainty of MST.

\subsection{Statistical uncertainty of MG}
Observe that $X=0$ means that reference MG and sample MG are equal; $X=1$ means that sample
MG is complement to reference MG. Let us pay attention that the latter situation for market graph is possible in only one
case, in contrast to MST.
For market graph one has $M_1 = \binom{N}{2} - M, M_2 = M$, where $N$ is the number of the vertices in the considered network
and $M$ is the number of edges in the given reference market graph. Since $M_1$ and $M_2$ are
constants, both measures of statistical uncertainty for MG are equivalent and can be defined from the equation:
$$
    \frac{1}{2}\sum\limits_{1 \leq i < j \leq N} \left[\frac{1}{\binom{N}{2} - M}P_n(x^{ij}_1 = 1) + \frac{1}{M}P_n(x^{ij}_2 = 1) \right] = \mathcal{E}_0.
$$

\begin{figure}[H]
    \centering
    \setlength\figureheight{10cm}
    \setlength\figurewidth{16cm}
    \includegraphics[scale=0.48]{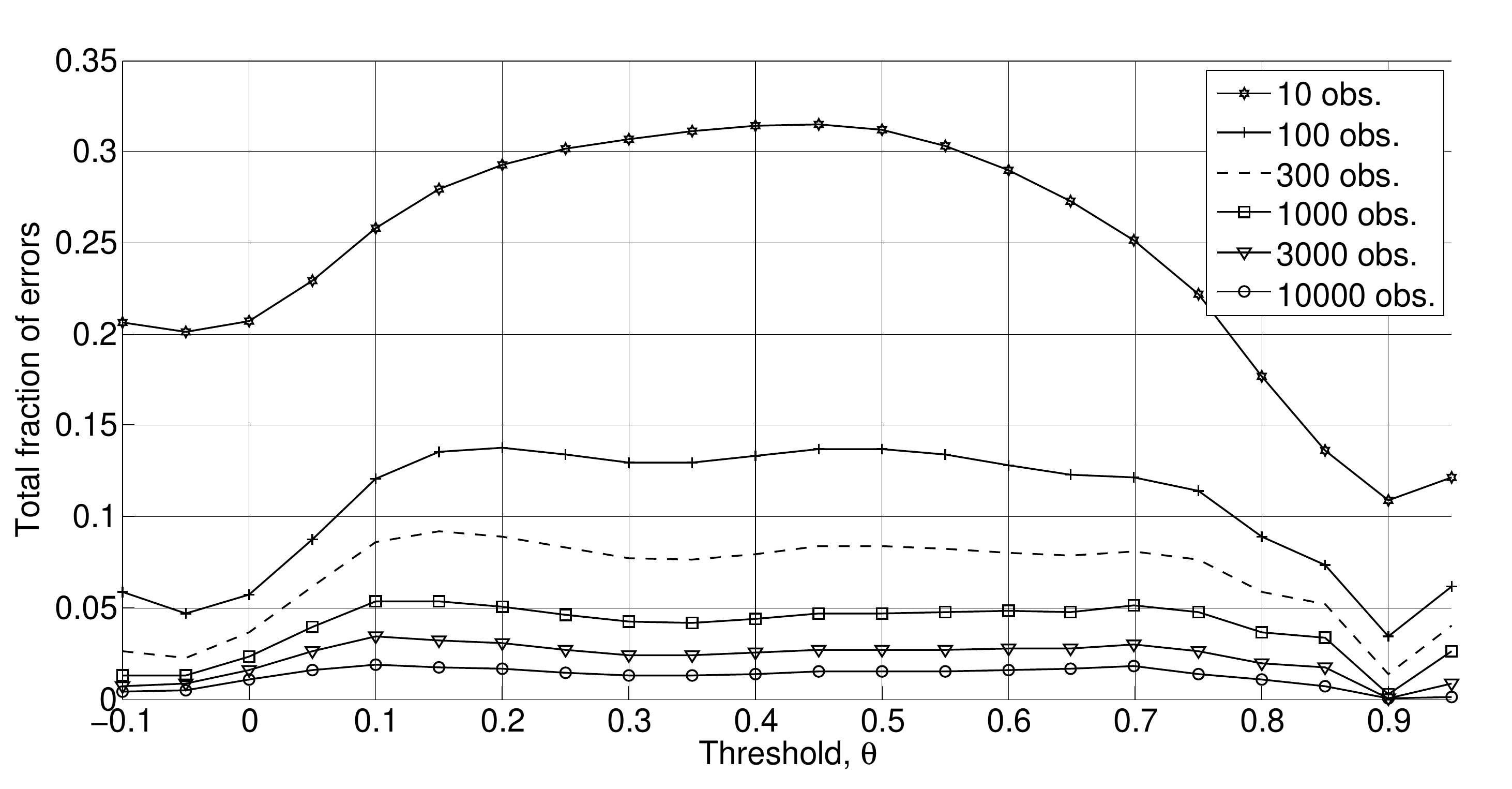}
    \caption{Total fraction of errors in market graphs.}
    \label{fig:graph_th_250}
\end{figure}

Results of the study of statistical uncertainty of MG are presented in Fig. \ref{fig:mst_250}, \ref{fig:graph_th_250}.
As one can see the condition $\mathcal{E}(\text{MG},n) \leq 0.1$ is achieved under the number of observed
periods $n_{\mathcal{E}} = 300$ for all thresholds $\theta \in [-0.1, 1]$, which is much more reasonable
than the statistical uncertainty of MST.

\subsection{Statistical uncertainty of MC}
Since a graph may contain many maximum cliques, in our experiments we choose the maximum clique with maximal weight (MCMW).
The weight of a clique in a market graph is the sum of weights of corresponding edges in the network. In each of all our
experiments there was only one maximum clique of maximal weight.
For maximum clique one has $M_1 = \binom{C_s}{2}$, $M_2=\binom{C_r}{2}$, where $C_s$ is a number of vertices in the sample MCMW and $C_r$ is a number of vertices in the reference MCMW.
Since $C_s$ and therefore, $M_1$ are random variables, the measure of statistical uncertainty for MCMW can be defined from the equation:
$$
    \frac{1}{2} E_n\left(\frac{X_1}{M_1}\right) + \frac{1}{2} \sum\limits_{1 \leq i < j \leq N}\frac{P_n(x^{ij}_2 = 1)}{M_2} = \mathcal{E}_0.
$$
Observe that $X=0$ means that reference MCMW and sample MCMW are equal.

$X=1$ means that $X_1=M_1$ and $X_2=M_2$.

$X_1=M_1$ corresponds to the situation when all edges of sample MCMW were included
incorrectly, i.e. vertices of sample MCMW induce the empty subgraph in the reference market graph.

$X_2=M_2$ corresponds to the situation when all edges of the reference MCMW are absent in a sample market graph, i.e. vertices of the reference MCMW induce empty subgraph in the sample market graph.

\begin{figure}[H]
    \centering
    \setlength\figureheight{10cm}
    \setlength\figurewidth{16cm}
    \includegraphics[scale=0.46]{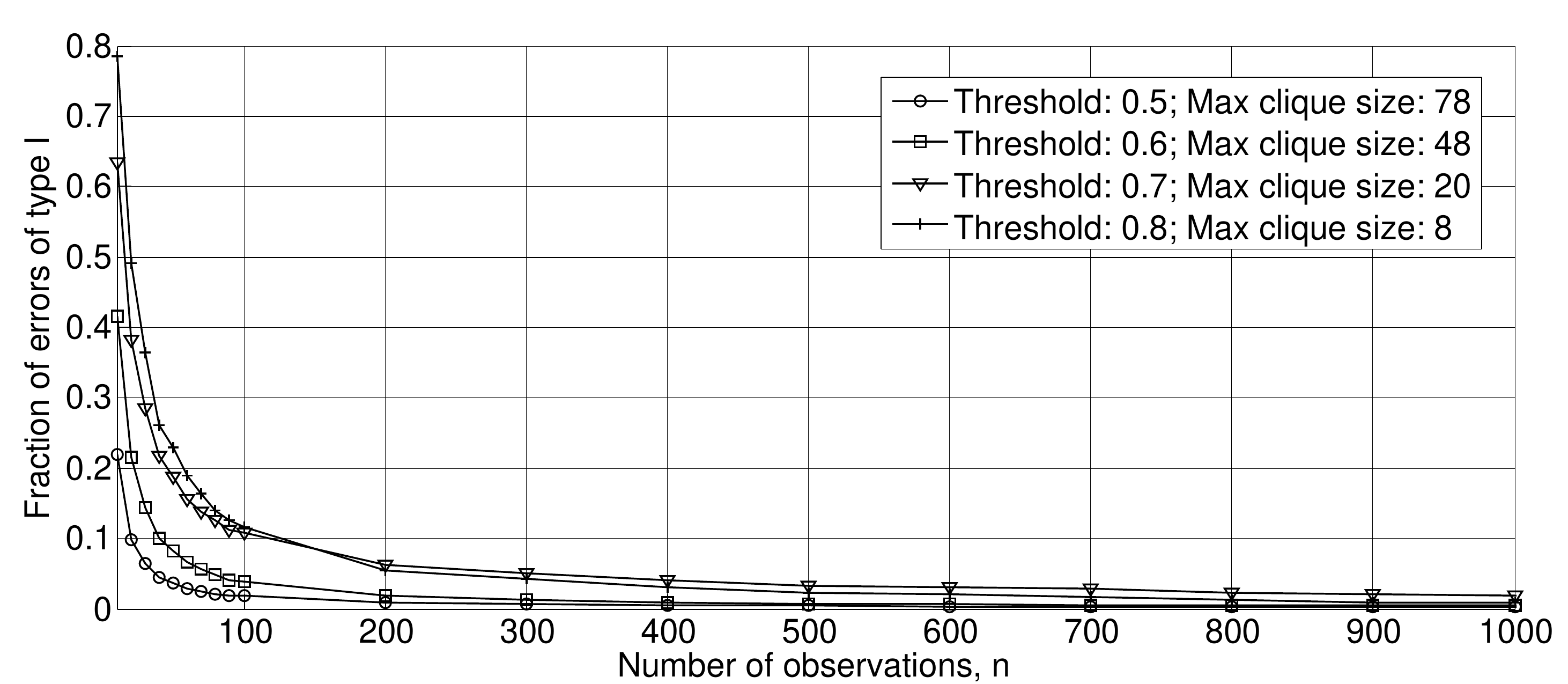}
    \caption{Fraction of errors of type I in cliques.}
    \label{fig:clq_max_w_A_250}
\end{figure}

\begin{figure}[H]
    \centering
    \setlength\figureheight{10cm}
    \setlength\figurewidth{16cm}
    \includegraphics[scale=0.46]{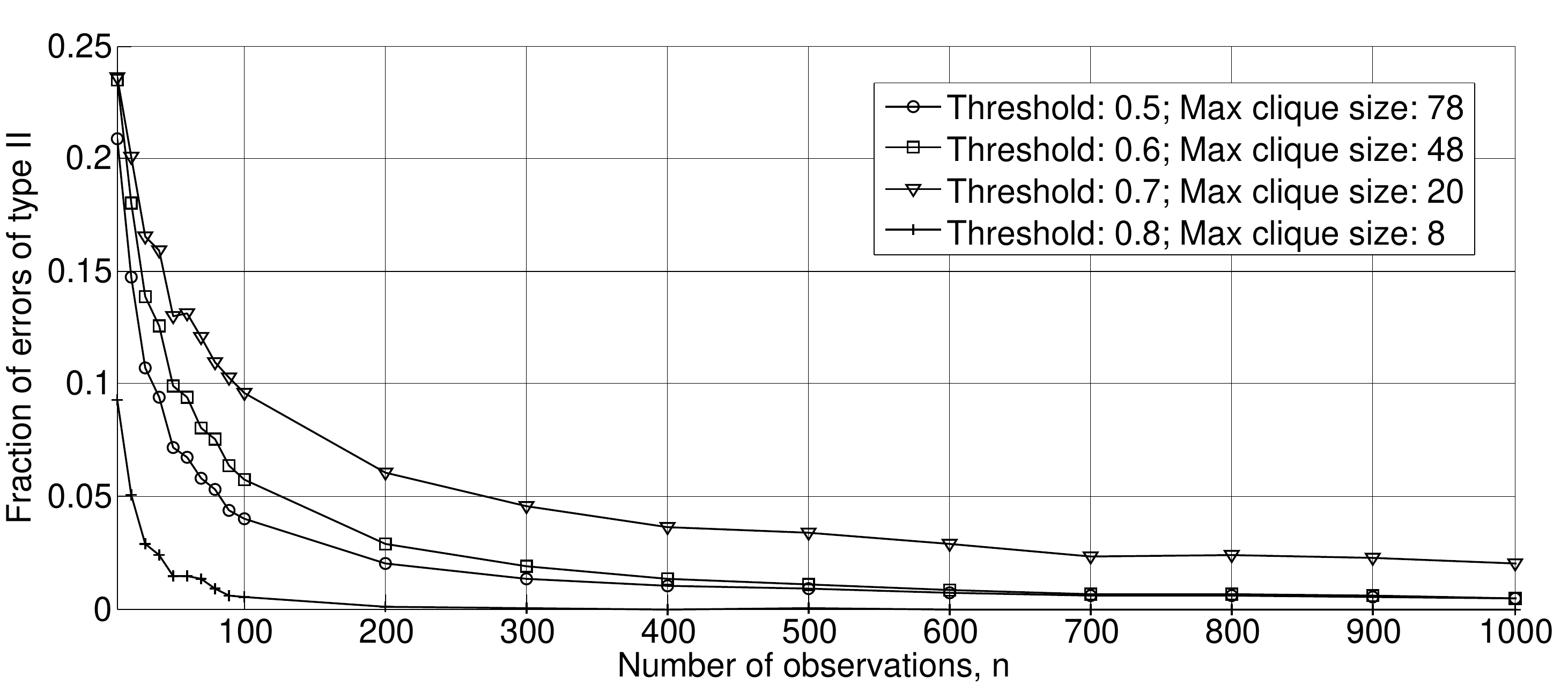}
    \caption{Fraction of errors of type II in cliques.}
    \label{fig:clq_max_w_B_250}
\end{figure}

\begin{figure}[H]
    \centering
    \setlength\figureheight{10cm}
    \setlength\figurewidth{16cm}
    \includegraphics[scale=0.46]{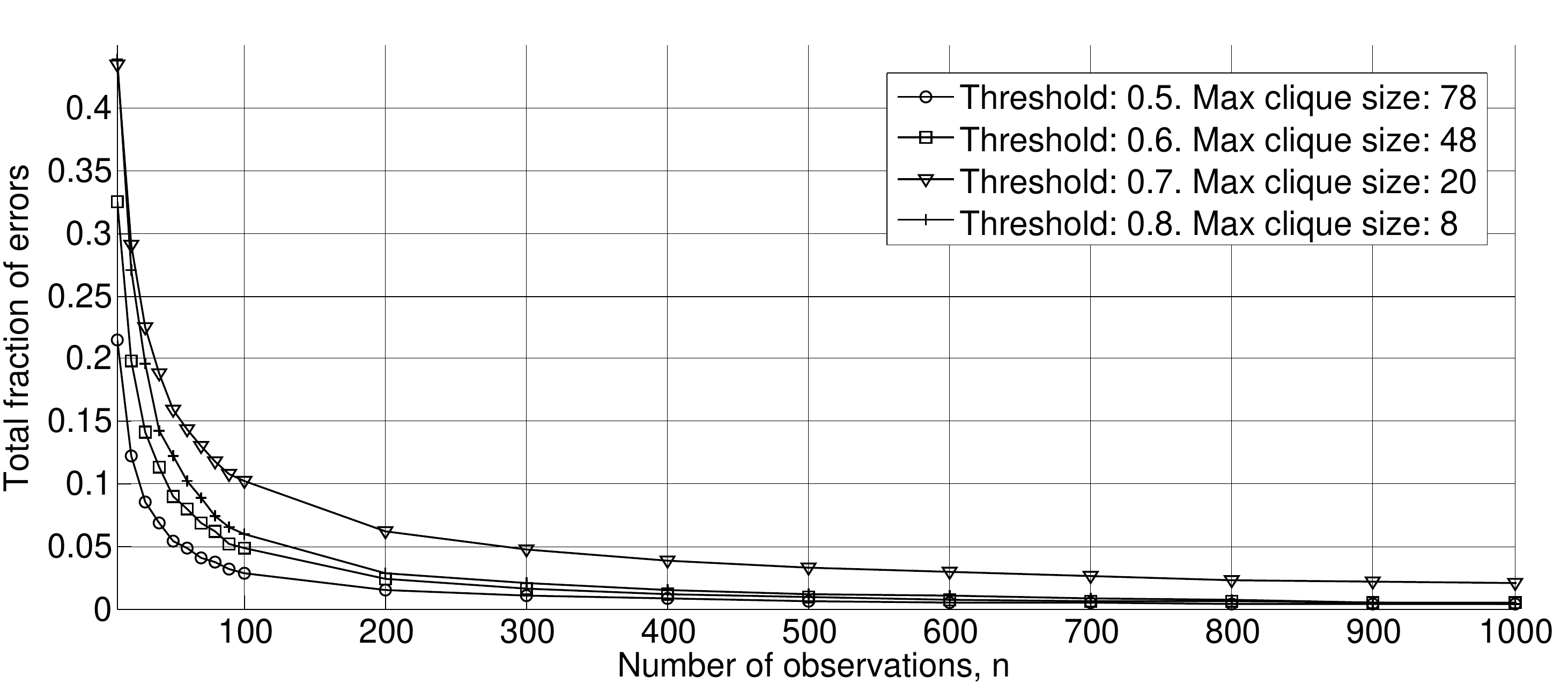}
    \caption{Total fraction of errors in cliques.}
    \label{fig:clq_max_w_250}
\end{figure}

Results of the study of statistical uncertainty of MCMW are presented in Fig. \ref{fig:clq_max_w_A_250}, \ref{fig:clq_max_w_B_250}, \ref{fig:clq_max_w_250}.
As one can see the condition $\mathcal{E}(\text{MCMW},n) \leq 0.1$ is achieved under the number of observed
periods $n_{\mathcal{E}} = 150$ for all considered thresholds $\theta$.
\\

\subsection{Statistical uncertainty of MIS}
Since a graph may contain many maximum independent sets, in our experiments we choose the maximum independent set with minimal weight (MISMW).
The weight of an independent set in a market graph is the sum of weights of corresponding edges in the network.
In each of our experiments there was only one maximum independent set of minimal weight.
For maximum independent set one has $M_1 = \binom{I_r}{2}$, $M_2=\binom{I_s}{2}$, where $I_r$ is a number of vertices in the reference MIS and $I_s$ is a number of vertices in a sample MISMW.
Since $I_s$ and therefore, $M_2$ are random variables, the measure of statistical uncertainty for MISMW is defined from the equation:
$$
    \frac{1}{2} \sum\limits_{1 \leq i < j \leq N}\frac{P_n(x^{ij}_1 = 1)}{M_1} + \frac{1}{2} E_n\left(\frac{X_2}{M_2}\right) = \mathcal{E}_0.
$$
Observe that $X=0$ means that reference MISMW and sample MISMW are equal.

$X=1$ means that $X_1=M_1$ and $X_2=M_2$.

$X_1=M_1$ corresponds to the situation when vertices of the reference MISMW induce complete subgraph in a sample market graph.

$X_2=M_2$  corresponds to the situation when vertices of a sample MISMW induce complete subgraph in the reference market graph.

\begin{figure}[H]
    \centering
    \setlength\figureheight{10cm}
    \setlength\figurewidth{16cm}
    \includegraphics[scale=0.46]{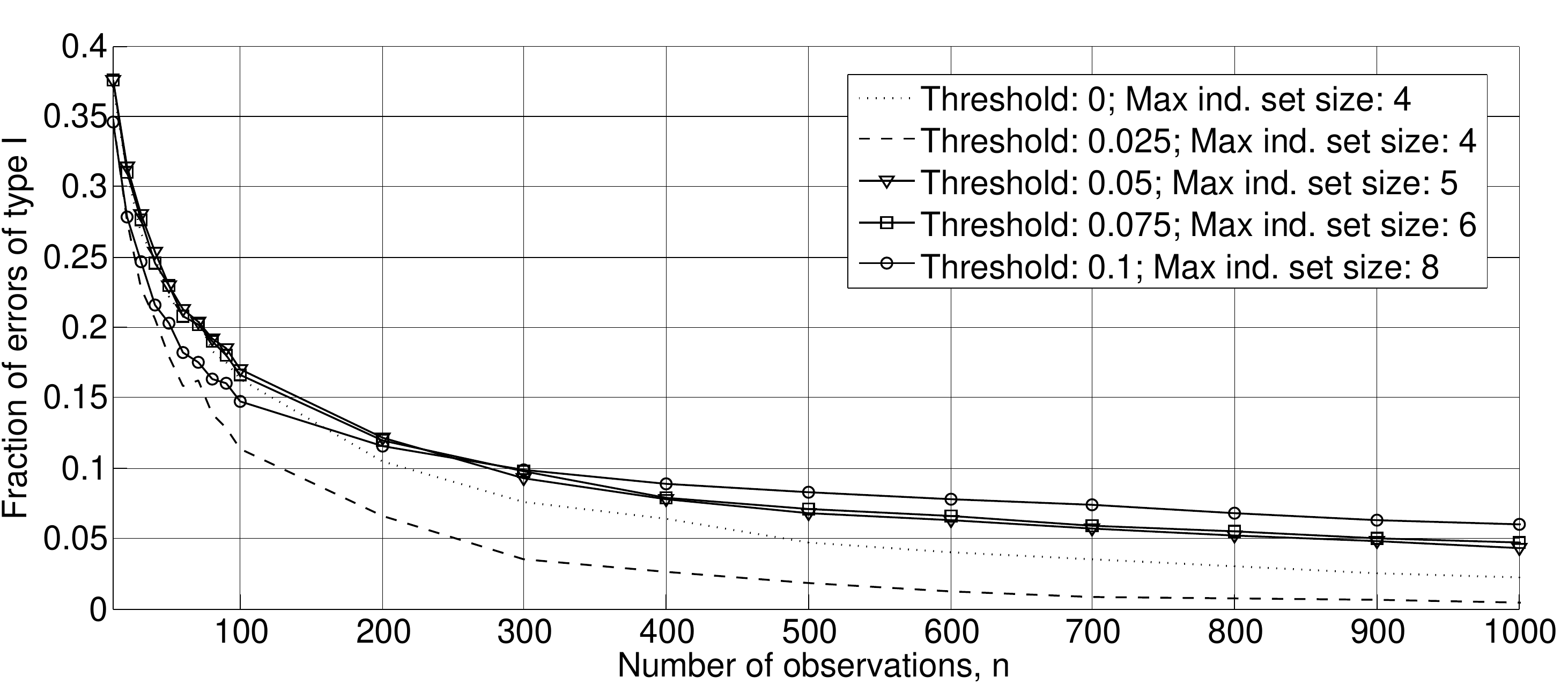}
    \caption{Fraction of errors of type I in an independent set.}
    \label{fig:is_min_w_A_250}
\end{figure}

\begin{figure}[H]
    \centering
    \setlength\figureheight{10cm}
    \setlength\figurewidth{16cm}
    \includegraphics[scale=0.46]{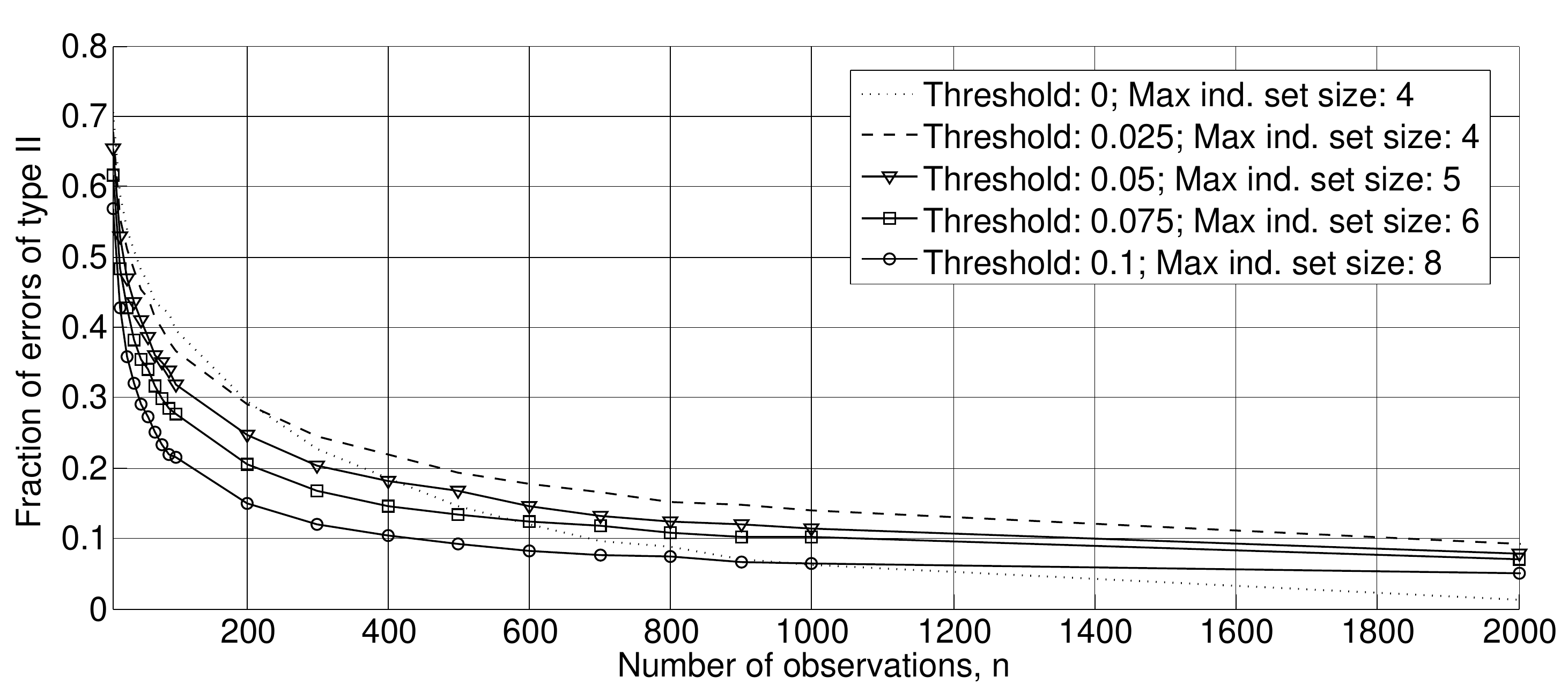}
    \caption{Fraction of errors of type II in an independent set.}
    \label{fig:is_min_w_B_250}
\end{figure}

\begin{figure}[H]
    \centering
    \setlength\figureheight{10cm}
    \setlength\figurewidth{16cm}
    \includegraphics[scale=0.46]{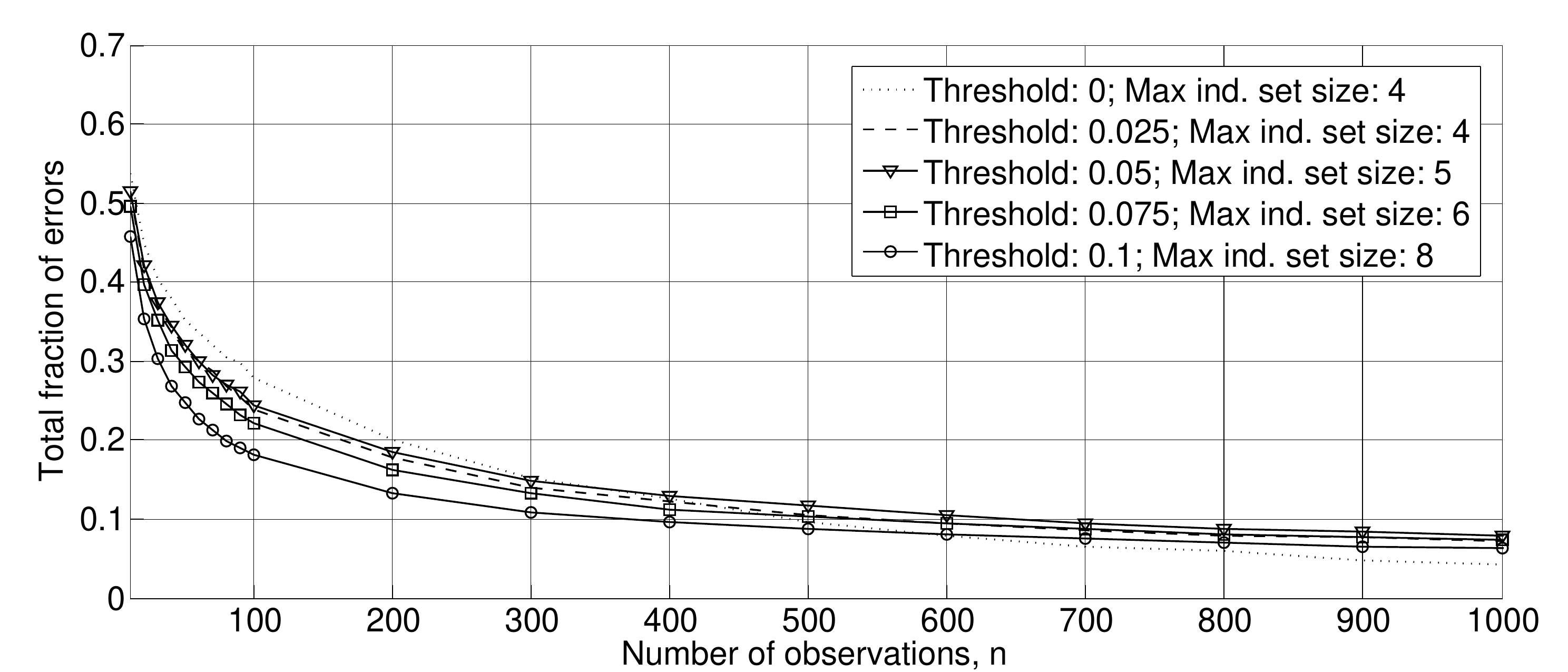}
    \caption{Total fraction of errors in an independent set.}
    \label{fig:is_min_w_250}
\end{figure}

Results of study of statistical uncertainty of MISMW are presented in Fig. \ref{fig:is_min_w_A_250},
\ref{fig:is_min_w_B_250}, \ref{fig:is_min_w_250}.
As one can see the condition $\mathcal{E}(\text{MISMW},n) \leq 0.1$ is achieved under the number of observed periods $n_{\mathcal{E}}$ = 700 for all considered thresholds $\theta$.
\\

\section{Note on hierarchical structure and the number of observations}\label{sec:note}
Hierarchical structure is an important characteristics in market network analysis \cite{Mantegna99, TAMM05}. 
In this note we discuss the stability of hierarchical structure obtained by MST filtering procedure. To evaluate the 
stability of hierarchical structure (topology) of MST we use the following invariant: vector of degrees of vertices 
ordered in ascending order. 
For MST from example 1 of section \ref{sec:NetMatAnalysis} this vector is $ (1, 1, 1, 1, 1, 1, 1, 1, 5, 5)$. 
We use $10 \times 10$ matrix from example 1 for the following numerical experiments.
\begin{enumerate}
    \item Simulate sample $x_{11}, \ldots, x_{1N}, \ldots, x_{n1}, \ldots, x_{nN}$.
    \item Calculate estimations $r_{ij}$ of parameters $\rho_{ij}$.
    \item In sample network (with weight matrix $||r_{ij}||$) find sample MST.
    \item Calculate vector of degrees for the sample MST.
    \item Repeat 1000 times steps 1-4 and calculate the frequencies of different vectors.
\end{enumerate}
The results are presented in the following table

\begin{center}
	\begin{table}[h]
		{\scriptsize
		\begin{tabular}{|c|c|c|c|c|c|c|c|}
			\hline
			Degree vec. / No. of observations &5 &10 &20 &100 &1000 &10000 &50000\\
			\hline
			(1, 1, 1, 1, 1, 1, 1, 1, 1, 9) &0 &0 &0 &0.004 &0 &0 &0\\
			\hline
			(1, 1, 1, 1, 1, 1, 1, 1, 2, 8) &0 &0 &0.002 &0.035 &0.04 &0 &0\\
			\hline
			(1, 1, 1, 1, 1, 1, 1, 1, 3, 7) &0 &0.001 &0.004 &0.065 &0.156 &0.032 &0.001\\
			\hline
			(1, 1, 1, 1, 1, 1, 1, 1, 4, 6) &0 &0 &0.005 &0.12 &0.349 &0.421 &0.352\\
			\hline
			\textbf{(1, 1, 1, 1, 1, 1, 1, 1, 5, 5)} &\textbf{0} &\textbf{0} &\textbf{0.005} &\textbf{0.075} &\textbf{0.248} &\textbf{0.545} &\textbf{0.647}\\
			\hline
			(1, 1, 1, 1, 1, 1, 1, 2, 2, 7) &0 &0 &0.011 &0.057 &0.01 &0 &0\\
			\hline
			(1, 1, 1, 1, 1, 1, 1, 2, 3, 6) &0 &0.005 &0.043 &0.166 &0.047 &0 &0\\
			\hline
			(1, 1, 1, 1, 1, 1, 1, 2, 4, 5) &0 &0.007 &0.041 &0.246 &0.15 &0.002 &0\\
			\hline
			(1, 1, 1, 1, 1, 1, 1, 3, 3, 5) &0 &0.007 &0.024 &0.016 &0 &0 &0\\
			\hline
			(1, 1, 1, 1, 1, 1, 1, 3, 4, 4) &0 &0.014 &0.017 &0.019 &0 &0 &0\\
			\hline
			(1, 1, 1, 1, 1, 1, 2, 2, 2, 6) &0 &0.014 &0.039 &0.035 &0 &0 &0\\
			\hline
			(1, 1, 1, 1, 1, 1, 2, 2, 3, 5) &0.004 &0.071 &0.151 &0.087 &0 &0 &0\\
			\hline
			(1, 1, 1, 1, 1, 1, 2, 2, 4, 4) &0.007 &0.042 &0.069 &0.044 &0 &0 &0\\
			\hline
			(1, 1, 1, 1, 1, 1, 2, 3, 3, 4) &0.017 &0.094 &0.142 &0.016 &0 &0 &0\\
			\hline
			(1, 1, 1, 1, 1, 1, 3, 3, 3, 3) &0.007 &0.005 &0.003 &0 &0 &0 &0\\
			\hline
			(1, 1, 1, 1, 1, 2, 2, 2, 2, 5) &0.006 &0.047 &0.046 &0.004 &0 &0 &0\\
			\hline
			(1, 1, 1, 1, 1, 2, 2, 2, 3, 4) &0.11 &0.262 &0.212 &0.01 &0 &0 &0\\
			\hline
			(1, 1, 1, 1, 1, 2, 2, 3, 3, 3) &0.137 &0.136 &0.084 &0.001 &0 &0 &0\\
			\hline
			(1, 1, 1, 1, 2, 2, 2, 2, 2, 4) &0.088 &0.06 &0.037 &0 &0 &0 &0\\
			\hline
			(1, 1, 1, 1, 2, 2, 2, 2, 3, 3) &0.394 &0.184 &0.048 &0 &0 &0 &0\\
			\hline
			(1, 1, 1, 2, 2, 2, 2, 2, 2, 3) &0.216 &0.049 &0.017 &0 &0 &0 &0\\
			\hline
			(1, 1, 2, 2, 2, 2, 2, 2, 2, 2) &0.014 &0.002 &0 &0 &0 &0 &0\\
			\hline
		\end{tabular}
		}
		\caption{Frequencies of degree vectors}
	\end{table}
\end{center}

The table shows that the true hierarchical structure (see Fig \ref{fig:ref_mst}) can be detected with probability 
more then  $1/2)$ only  in the case of more then 10000 observations and this probability is increasing very slowly.

Finally we would like to make a rather obvious
remark that there is a lower bound on the number of observation
periods (sample size) $n$ for the construction of $N$-vertex
sample network. Namely, the number of observed values $x_{tk},
t=1, \ldots, n, k=1,\dots, N$ should be at least as much as the
number of estimated parameters $\rho_{ij} =
\sigma_{ij}/\sqrt{\sigma_{ii}\sigma_{jj}}, 1\leq i < j \leq N$, in
order to eliminate a functional dependence between estimations
$r_{ij} = s_{ij}/\sqrt{s_{ii}s_{jj}}$ of parameters $\rho_{ij}$.
Total number of observed values $x_{tk}, t=1, \ldots, n,
k=1,\dots, N$ during $n$ days is $nN$. We calculate $N(N-1)/2$
estimations $r_{ij}$ of unknown parameters $\rho_{ij}$ using $nN$
observed values. Therefore, if $n < (N-1)/2$, then  arise
unavoidable functional dependence between $r_{ij}$ regardless of
the existence of such relationships between $\rho_{ij}$. This
dependence introduces false relationship between stock returns. The case were $n<N$ is of practical importance and recently it attracted some attention in the literature (see \cite{Hub} where the partial correlations are used at the place of Pearson correlations). 

\begin{figure}[H]
    \centering
    \setlength\figureheight{10cm}
    \setlength\figurewidth{16cm}
    \includegraphics[scale=0.41]{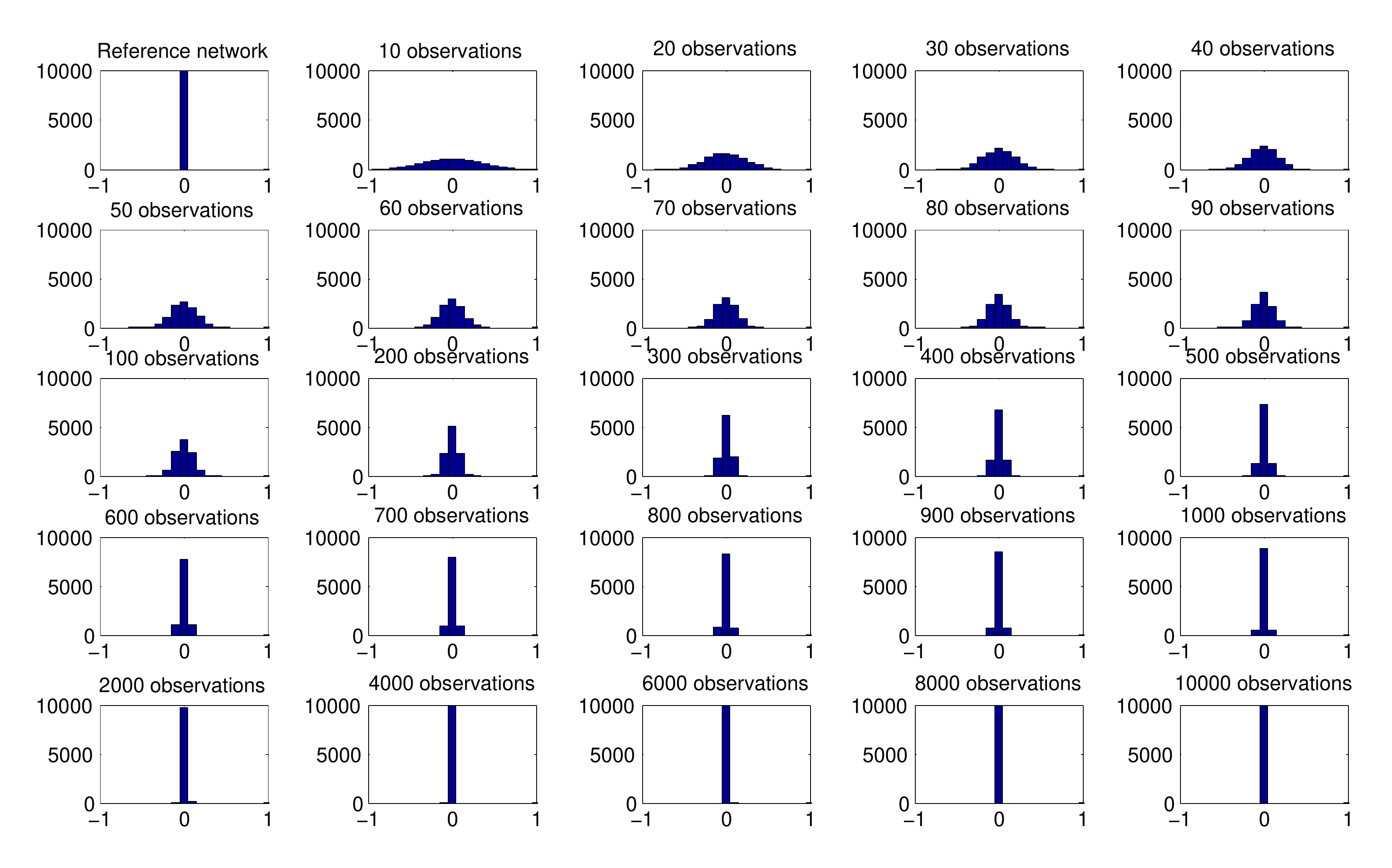}
    \caption{Distribution of correlation coefficients depending on a number of observations.}
    \label{fig:hist_model}
\end{figure}

\noindent\textbf{Example 2} \\
We illustrate this fact by the following example. Let us consider a reference network
with identity $100 \times 100$ matrix $||\rho_{ij}||$,
which corresponds to a market
with 100 pairwise independent stocks. Distribution of edge weights of network can be clearly represented by histogram. The histograms of edge weights for reference network and for $n$-periods sample networks for different values of $n$ are presented in Fig. \ref{fig:hist_model}.
As one can see for small $n$ the histograms for $n$-period sample networks significantly differ from
the reference network histogram.
Note that for $n < 50$ this difference is not only due to random nature of observations, but also because of emergence of significant number of false relationships between stock returns which are absent for $n \geq 50$.

\begin{figure}[H]
    \centering
    \setlength\figureheight{10cm}
    \setlength\figurewidth{16cm}
    \includegraphics[scale=0.41]{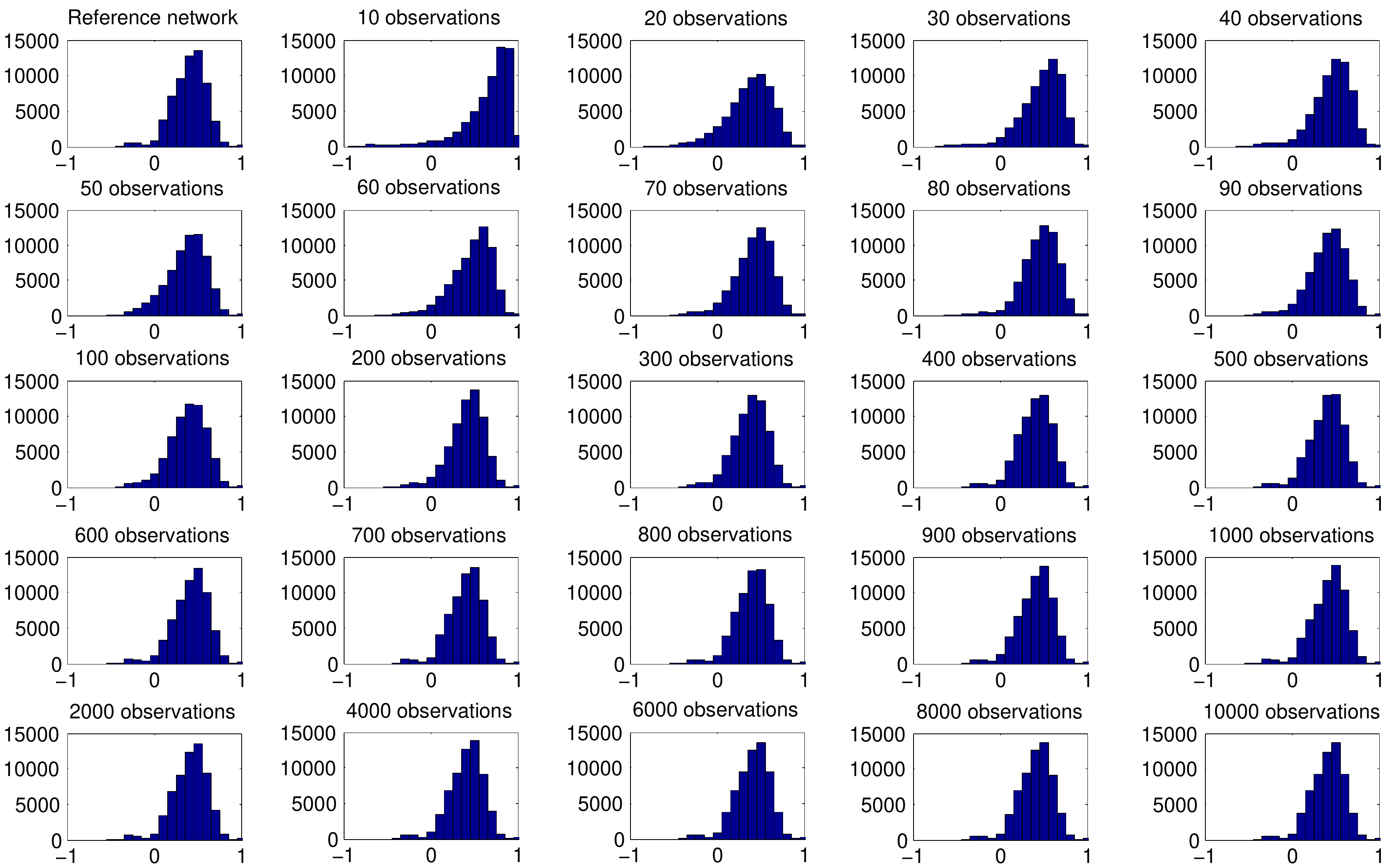}
    \caption{Histograms of edge weights for US market true and sample networks.}
    \label{fig:hist_250}
\end{figure}

Similar historgram for the US reference network is presented in Fig. \ref{fig:hist_250}.

\section{Conclusions}\label{sec:Conclusions}
The main results of the present paper are introduction and application of two measures of statistical uncertainty of market network analysis.
One of the proposed measure of statistical uncertainty is based on conditional risk known as fundamental notion of
Wald's statistical decision theory. Another measure of statistical uncertainty is based on average fraction of errors.
In both cases the  measure of statistical uncertainty is the number of observations needed to obtain results with given
confidence.
It is shown that for some structures the second measure is a special case of the first one.

We apply our approach to analyse the statistical uncertainty of US stock market. Our experimental study shows that
Market Graph, Maximum Clique, Maximum Independent Set are more reliable with respect to statistical uncertainty than
Minimum Spanning Tree and Planar Maximally Filtered Graph.
In particular, for 250 stocks of US market to obtain 90\% of confidence for minimum spanning tree one needs at
least 10000 observations.
The same confidence for Market Graph is already reached with 300 observations; for Maximum Clique -- 150 observations
and for Maximum Independent Set -- 700 observations.


\end{document}